\documentclass[12pt,halfline,a4paper]{ouparticle}
\usepackage{xcolor}
\usepackage{hyperref}
\hypersetup{
  colorlinks,
  citecolor=blue,
  linkcolor=blue,
  urlcolor=black}

\usepackage{booktabs}
\usepackage{graphicx}
\usepackage{dcolumn}
\usepackage{float}
\usepackage{bm}

\begin{document}

\title{\LARGE\textbf{Ni/Bi bilayers: The effect of thickness on the superconducting properties}}

\author{\normalsize
Gabriel Sant'ana \footnote{gabriel.santana.dasilva@outlook.com}, David Möckli, Alexandre da Cas Viegas, Paulo Pureur and Milton A. Tumelero \footnote{matumelero@if.ufrgs.br}\\
\small\textit{Instituto de F\'{i}sica, Universidade Federal do Rio Grande do Sul, 91501-970 Porto Alegre, Brazil}\\
}
\date{}
\vspace{1cm}
\abstract{ Nickel/Bismuth (Ni/Bi) bilayers have recently attracted attention due to the occurrence of time-reversal symmetry breaking in the superconducting state. Here, we report on the structural, magnetic and electric characterization of thin film Ni/Bi bilayers with several Bi thicknesses. We observed the formation of a complex layered structure depending on the Bi thickness caused by the inter-diffusion of Bi and Ni which leads to the stabilization of NiBi$_{3}$ at the Bi/Ni interface. The superconducting transition temperature and the transition width are highly dependent on the Bi thickness and the layer structure. Magnetoelectric transport measurements in perpendicular and parallel magnetic fields were used to investigate the temperature-dependent upper critical field within the framework of the anisotropic Ginzburg-Landau theory and the Werthamer–Helfand–Hohenberg model. For thicker samples, we observed a conventional behavior, similar to that shown by NiBi$_{3}$ bulk samples, including a small Maki parameter ($\alpha_{M}$ =  0), no spin-orbit scattering ($\lambda_{SO}$= 0) and nearly isotropic coherence length ($\gamma$ = $\xi_{\perp}$(0)/$\xi_{||}$(0) $\approx$ 1). The values obtained for these properties are close to those characterizing NiBi$_{3}$ single crystals. 
On the other hand, in very thin samples the Maki parameter increases to about $\alpha_{M}$ = 2.8. In addition, the coherence length becomes anisotropic ($\gamma$ = 0.32) and spin-orbit scattering ($\lambda_{SO}$= 1.2) must be taken into account.  Our results unequivocally show that the properties characterizing the superconducting state in the Ni/Bi are strongly dependent on the sample thickness.}

\keywords{Ni/Bi bilayer, NiBi$_{3}$ compound; spin-orbit superconductors; unconventional superconductivity}
\maketitle

\section{\label{sec:level1}Introduction}

Time reversal symmetry-breaking (TRSB) in the superconducting state has been a strongly discussed topic in condensed matter physics. Superconductors showing TRSB have been considered as a great promise for technological advancements in quantum computing and spintronics, as they are expected to exhibit Majorana edge modes \cite{wilczek2009majorana,nayak2008non} and spin-triplet pairing \cite{ran2019nearly}. 
Currently, this phenomenon has been observed in uranium-based heavy fermions and in the perovskite Sr$_{2}$RuO$_{4}$ \cite{saxena2000superconductivity, aoki2001coexistence,luke1998time}. However, the occurrence of spontaneous TRSB superconductors in non-centrosymmetric 2D systems had not been reported until Gong \textit{et al.} observed this property in nickel/bismuth (Ni/Bi) bilayers \cite{gong2017time}.

Interestingly, while Ni does not exhibit superconductivity (SC) and Bi only displays this property below 0.5 mK at ambient pressure \cite{prakash2017evidence}, a superconducting transition with a critical temperature ($T_c$) of approximately 4 K occurs when both elements are grown as a Ni/Bi thin film bilayer. 
This heterostructure was initially investigated by Moodera \textit{et al.} \cite{moodera1990superconducting} and LeClair \textit{et al.} \cite{leclair2005coexistence}, who used tunneling measurements to suggest that SC potentially occurs across the entire bilayer. Similar conclusions were later reported in the experimental studies of Gong \textit{et al.}  \cite{gong2017time} and Chauhan \textit{et al.}  \cite{chauhan2019nodeless}, the latter arguing in favor of a $p$-wave nature of symmetry pairing.  Other experimental works have observed signatures of unconventional SC consistent with chiral $p$-wave behavior \cite{gong2015possible,wang2017anomalous}.

Although many studies consider the Ni/Bi bilayer as a potential unconventional superconductor, the presence of a NiBi$_{3}$ layer formed 
at the interface may be an alternative explanation for the observed phenomenology. 
This particular intermetallic compound is described 
as a strong coupling $s$-wave superconductor with $T_c$ around 4.1 K \cite{zhao2018singlet,zhu2012surface,fujimori2000superconducting}, essentially the same $T_c$ as for the Ni/Bi bilayer heterostructure. 
Regardless of the preparation technique, spontaneous formation of NiBi$_3$ is observed when the deposition temperature of the bilayer is higher than 110K \cite{siva2015spontaneous,vaughan2020origin,liu2018superconductivity}. 
However, when the Ni/Bi bilayer is prepared at 4.2 K, to avoid interdiffusion, Liu \textit{et al.} \cite{liu2018superconductivity} verified that neither NiBi$_3$ is formed nor the resulting bilayer is superconducting.


In essence, the presence of an interfacial NiBi$_{3}$ layer appears to be a necessary condition for the observation of a superconducting transition in the Ni/Bi bilayer. 
A theoretical model proposed by Chao considers the presence of NiBi$_{3}$ acting as a mechanism to induce SC in the Bi layer via the proximity effect \cite{chao2019superconductivity}. 
In addition, the model takes into account the strong spin-orbit interaction related to Bi and the exchange coupling coming from Ni to provide a reasonable explanation for the $p$-wave-like Andreev reflection signatures seen previously \cite{gong2015possible}. 
In other words, the model proposes a chiral $p$-wave symmetry with non-trivial topological phase to describe the superconducting state in the Bi/Ni bilayer system. Therefore, at this point, it seems unclear whether the SC in Bi/Ni originates directly from NiBi$_{3}$ at the interface or whether it is induced by the proximity effect in the adjacent Bi layer. 
It is nonetheless unclear if these configurations host an unconventional superconducting state.

To further contribute, we study the layered Bi/Ni system with its superconducting properties including: 
critical temperature, upper critical field, and coherence length as functions of the Bi layer thickness.
Our analyses are based on Ginzburg-Landau and Werthamer–Helfand–Hohenberg (WHH) models, considering two different magnetic field configurations: parallel (aligned with the current) and perpendicular (out of the film plane). 
Our findings suggest the presence of a thickness-dependent crossover point at which unconventional properties potentially emerge.

\section{\label{sec:level2}Experimental details}

We grew Ni/Bi bilayer thin films on silicon oxide (100 nm) covered silicon wafer substrates by using the magnetron sputtering deposition technique. 
The substrates were kept at room temperature. 
We prepared several samples with a fixed 8 nm layer of Ni and layers of Bi with thicknesses ranging from 10 to 80 nm. 
The purity of both Ni and Bi targets was 99.99$\%$ and the power employed for the depositions was 65 W (RF) for the Ni target and 10 W (DC) for the Bi target. 
A capping Ti layer of 2 nm was applied to all samples to prevent oxidation. 
The vacuum chamber base pressure was maintained at 2$\times$10$^{-7}$ Torr, and the deposition working pressure was set to 1$\times 10^{-3}$ Torr by using 200 sccm flux of ultrapure argon gas (99.9999$\%$).

Rutherford backscattering spectroscopy (RBS) using 1.5 MeV He+ beam was employed to determine the samples' thicknesses. 
To study the crystallographic properties we applied X-ray diffraction (XRD) in the $\Theta - 2\Theta$ configuration with Cu$\alpha$1 radiation. 
Magnetization was measured as a function of the field at room temperature in a VSM EV9 magnetometer manufactured by Microsense Inc. 

For the magneto-electrical measurements, we implemented a standard four-probe method where the in-plane current was applied at the extremities of the sample, while two inner contacts were used to measure the longitudinal voltage drop. 
In this arrangement, two magnetic field configurations were used: (i) parallel, with the field applied along the current direction, and (ii) perpendicular to the substrate plane. The magneto-transport experiments were performed using a set-up based on a CMag9 cryocooler manufactured by Cryomagnetics Inc. 

\section{\label{sec:level3}Results and Discussion}

\subsection{
Structural and magnetic characterization}\label{sec:characterization}

\begin{figure}[h]
    \centering
    \includegraphics[height=13.5cm, width=8.5cm]
    {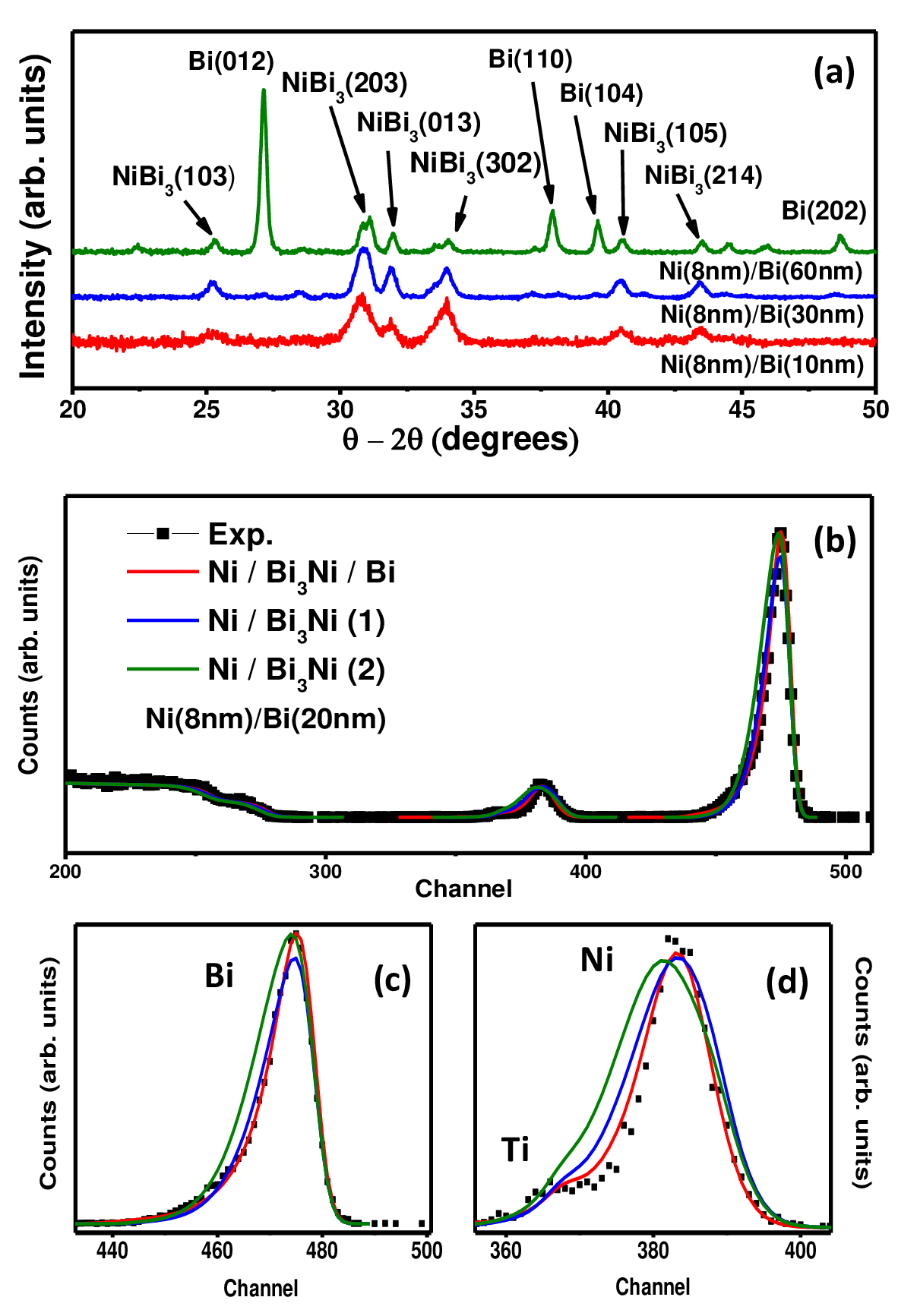}
    \caption{(a) The XRD pattern of Ni(8nm)/Bi$(x)$, for $x=10$ nm (red), $x=30$ nm (blue) and $x=60$ nm (green). (b) 
    The RBS spectra for Ni(8nm)/Bi(20nm) with close-ups for the 
    (c) Bi and (d) Ni and Ti peaks. The green and blue lines in panels (c) and (d) correspond to different SIMNRA simulations for a NiBi$_{3}$/Ni bilayer, while the red line is a simulation for a Ni/NiBi$_{3}$/Bi trilayer. }
    \label{fig:structural_charact}
\end{figure}

Fig. \ref{fig:structural_charact}$\color{blue}\text{a}\color{black}$ displays the XRD patterns obtained for Ni(8nm)/Bi($x$) samples with $x$ = 10, 30 and 60 nm. 
The presence of NiBi$_{3}$ peaks is evident, while no peaks of individual Bi and Ni layers were observed. This result is in agreement with previous reports that also notice the formation of the inter-metallic NiBi$_{3}$ compound \cite{siva2015spontaneous,liu2018superconductivity,vaughan2020origin}. 
The XRD pattern is consistent with the orthorhombic structure of NiBi$_{3}$ with $Pnma$ space group. 
We did not observe the NiBi compound
in our samples, which contrasts with a previous study \cite{liu2018superconductivity}.

Additional information on the structural properties 
was obtained from Gaussian fitting the XRD peaks in Fig. \ref{fig:structural_charact}$\color{blue}\text{a}\color{black}$. 
The full-width at half maxima (FWHM) decreases with growing sample thickness, indicating that thicker samples
are more crystalline. 
In the Ni(8nm)/Bi(60nm) sample (shown in green), the observed pure Bi peaks (012), (110), (104), and (202) are consistent with the R-3m trigonal structure. 
The presence of solid Bi suggests a limit in the formation of NiBi$_{3}$, which can be interpreted based on two contributions: 
(i) A diffusive limited process, in which the Bi diffuses into the Ni layer forming the NiBi$_{3}$ layer; 
(ii) A kinetic limited process, in which the Bi reacts with Ni forming the NiBi$_{3}$ layer until one of the layers (Bi or Ni) is fully consumed.

Fig. \ref{fig:structural_charact}$\color{blue}\text{b}\color{black}$ presents the RBS spectrum obtained for the Ni(8nm)/Bi(20nm) sample, whereas Figs. \ref{fig:structural_charact}$\color{blue}\text{c,d}\color{black}$ provide zoomed-in views of the Bi and Ni peaks, respectively. The capping Ti layer is also detected as a shoulder in Fig. \ref{fig:structural_charact}$\color{blue}\text{d}\color{black}$. To determine the sample stoichiometry and thickness, we utilized the SIMNRA software and the values of bulk mass density. In Figs. \ref{fig:structural_charact}$\color{blue}\text{c,d}\color{black}$, the green and blue lines correspond to simulations of a Ni/NiBi$_{3}$ bilayer with: 
(1) focus on simulating the maxima of the Ni peak; and (2) focus on simulating the maxima of the Bi peak, whereas the red line corresponds to the simulation of the trilayer Ni/NiBi$_{3}$/Bi. 
The latter offers the best fit for the experimental data. 
This suggests that even for the bilayer with the lowest Bi thickness, a thin layer of Bi remains after the formation of NiBi$_{3}$ at the interface. 
Consequently, this observation can be assumed as evidence in favor of the diffusive limited NiBi$_{3}$ formation process. Additionally, the RBS simulation for the Ni(8nm)/Bi(20nm) indicates a thickness of 11 nm for Bi, 18 nm for NiBi$_{3}$ and 7 nm for Ni, approximately. This means that the relative consumption rate between Ni and Bi is close to (1:9) nm, which agrees with the findings of Vaughan \textit{et. al} \cite{vaughan2020origin}, who makes use of a polarized neutron scattering technique.

\begin{figure}[h]
    \centering
    \includegraphics[height=7.5cm, width=7.5cm]{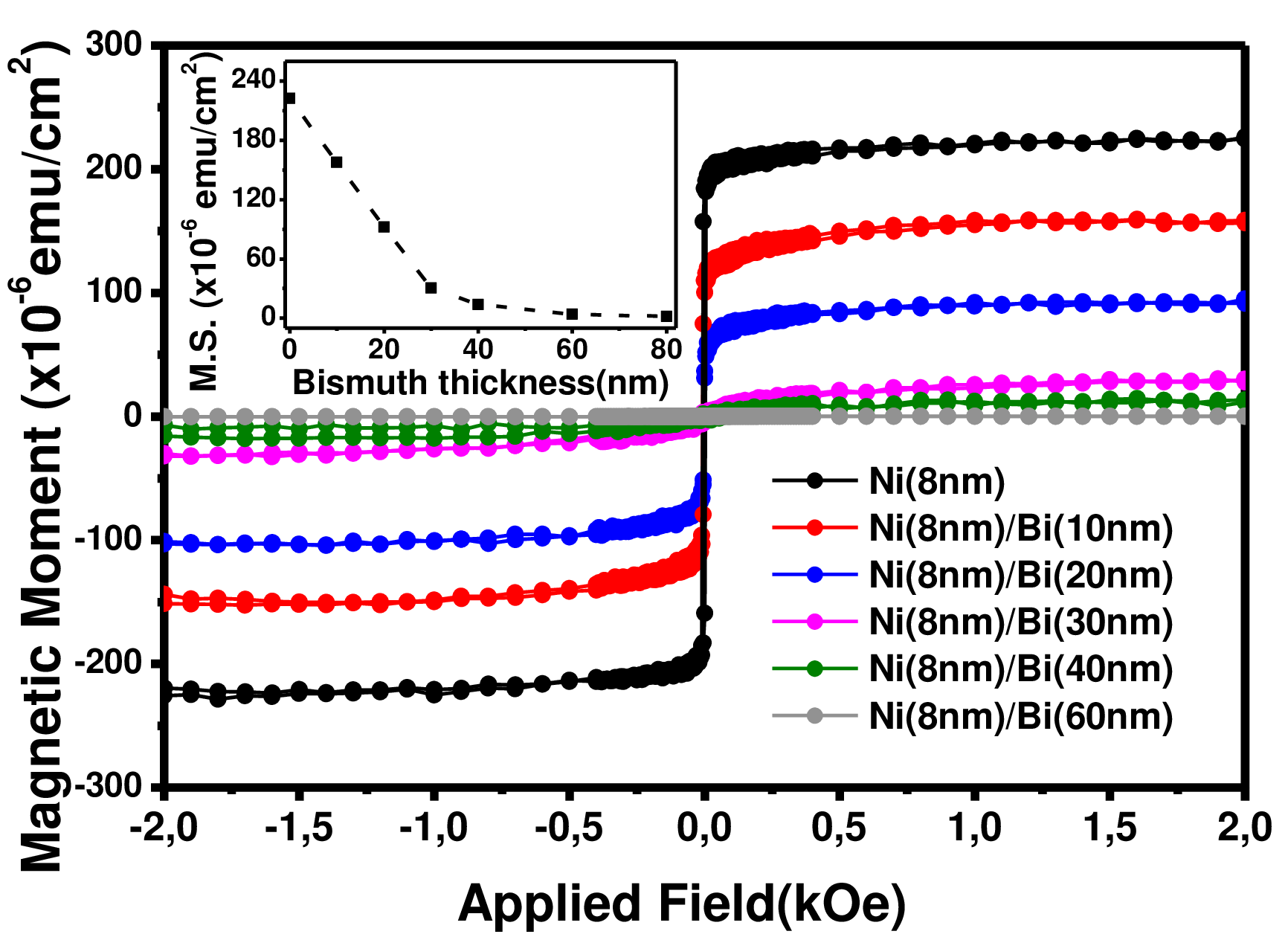}
    \caption{Magnetic moment normalized by the area of each sample. Measurements were made at room temperature as functions of the in-plane  magnetic field. Inset: magnetic moment of saturation (M.S.) normalized by area vs bismuth thickness.}
    \label{fig:magnetization}
\end{figure}

Fig. \ref{fig:magnetization} shows the magnetic moment normalized by the sample's area as a function of the in-plane applied magnetic field. 
These experiments were carried out at room temperature. Previous studies have not reported any magnetic order in NiBi$_{3}$ \cite{kumar2011physical,silva2013superconductivity}. 
Therefore, we assume that the magnetic moment observed in our samples originates from the Ni layer. 
Results in Fig. \ref{fig:magnetization} indicate that the presence of the Bi layer with increasing thickness gradually suppresses the Ni magnetic moment. 
The inset in Fig. \ref{fig:magnetization} depicts the magnetic moment of saturation (M.S.) for different Bi thicknesses. Clearly, the M.S. vanishes for samples where the Bi layers have a thickness larger than 40 nm. 
The formation of the NiBi$_{3}$ compound at the interface of the bilayers is probably the most important effect leading to the progressive suppression of ferromagnetism.

We estimated the thickness of the ferromagnetic portion of the Ni layer by assuming the bulk magnetization and structural parameters. 
Interestingly, for an 8 nm pure Ni layer (black dots in Fig. \ref{fig:magnetization}) the estimation yielded an effective ferromagnetic Ni layer thickness of about 4.8 nm. 
This suggests the occurrence of magnetically dead Ni sub-layers within the thin film structure. 
Such an interpretation was previously proposed by Liebermann \textit{et. al} to account for similar observations in Ni thin films at room temperature \cite{liebermann1970dead}. 
In our case, the dead layers might be related to the oxidation of Ni close to the interface with the SiO$_2$ substrate. Regarding the Ni/Bi bilayers, we estimate thicknesses for the ferromagnetic Ni of 3.4 nm, 2.1 nm, and 0.8 nm for Ni(8nm)/Bi($x$) with $x$ = 10, 20 and 30 nm, respectively. These results indicate a consistent consumption rate close to (1:9) nm, similar to the one obtained from the RBS analysis.

Based on the characterization, we are led to the conclusion that two distinct types of Ni/Bi layered structures are present in our bilayers. 
For Bi depositions up to 30 nm, we effectively obtained a Ni/NiBi$_{3}$/Bi trilayer, resembling a heterostructure configuration of metal/superconductor/ferromagnet. 
In contrast, Bi thicknesses exceeding 30 nm reveal the occurrence of a progressive suppression of ferromagnetism in the Ni film, so that the samples behave as a NiBi$_{3}$/Bi bilayer, which is akin to a metal/superconductor system. 
Here, we consider that the Bi layer behaves as a metallic component, which is expected due to its low thickness \cite{xiao2012bi}.

\subsection{Thickness-dependence of the critical temperature}

\begin{figure}[h]
    \centering
    \includegraphics[height=11.3cm, width=7.5cm]{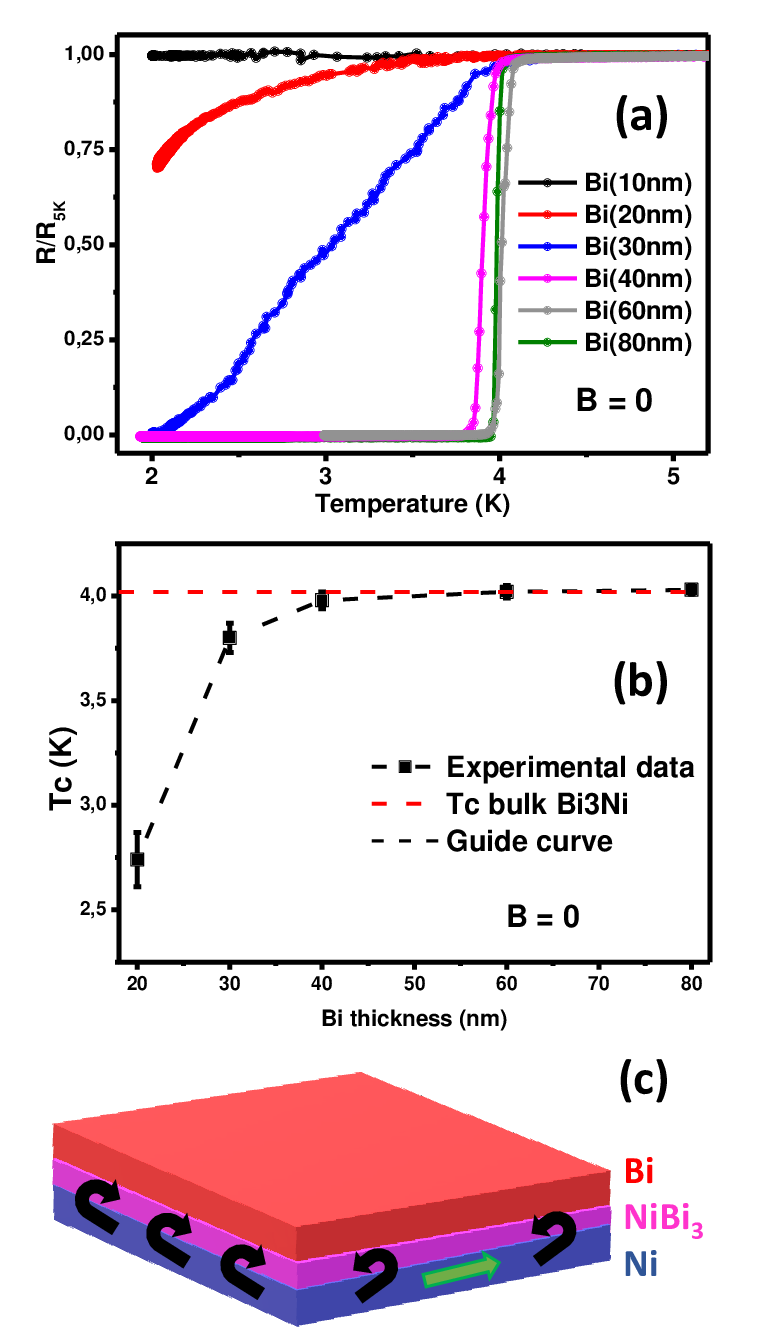}
    \caption{(a) Normalized resistance as a function of temperature for Ni(8nm)/Bi($x$) samples.
    (b) $T_c$ as a function of Bi thickness. 
    (c) Schematic illustration of the Ni/Bi$_3$Ni/Bi trilayer. The black arrows indicate the stray field originating from the Ni layer and green arrow corresponds to the in-plane magnetization. 
    }
    \label{fig:Tc_thickness}
\end{figure}

Fig. \ref{fig:Tc_thickness}$\color{blue}\text{a}\color{black}$  shows the normalized resistance ($R/R_{5K}$) vs. temperature, while Fig. \ref{fig:Tc_thickness}$\color{blue}\text{b}\color{black}$ displays $T_c$ as a function of the Bi thickness. 
Since our lowest reachable temperature is 1.8 K,
we define $T_c$ as the temperature where the measured resistance decreases to 90\% of the normal state resistance. 
Similar profiles of $T_c$ vs. thickness were reported in Refs. \cite{gong2017time,gong2015possible,vaughan2020origin,liu2018superconductivity} for Ni/Bi bilayers.  
For samples deposited with higher Bi thickness, i.e. Ni(8nm)/Bi(40nm), Ni(8nm)/Bi(60nm) and Ni(8nm)/Bi(80nm), the superconducting transition is sharp and has a $T_c \approx 4.05$K, similar to those reported for the NiBi$_{3}$ compound (red dashed-line on Fig. \ref{fig:Tc_thickness}$\color{blue}\text{b}\color{black}$). 
On the other hand, a large broadening in the superconducting transition is observed for the samples Ni(8nm)/Bi(20nm) and Ni(8nm)/Bi(30nm). The transition for the sample Ni(8nm)/Bi(10nm) was not observed to occur in the measured temperature range. The broadening of the superconducting transition is usually related to the disorder. 
We have indeed observed some increase in lattice disorder by looking at the FWHM in the XRD patterns for samples with small Bi thicknesses. Nonetheless, we would like to point out that the transition widens quite significantly when the Bi thickness is reduced below the 40 nm threshold. 
For instance, the transition broadening in the Ni(8nm)/Bi(30nm) sample is about 10 times larger than that in the Ni(8nm)/Bi(40nm) specimen. The FWHM, however, changes much more slowly in the same range of Bi thickness. 
Therefore, we believe that the broad transition might be related to other effects apart from lattice disorder.

By comparing the $T_c$ curve in Fig. \ref{fig:Tc_thickness}$\color{blue}\text{b}\color{black}$ with the inset in Fig. \ref{fig:magnetization}, we observe that a strong widening of the superconducting transition occurs in samples with a non-zero magnetic moment. 
This leads us to speculate that the remaining ferromagnetic Ni layer generates a highly non-uniform magnetic stray field, as depicted in Figure \ref{fig:Tc_thickness}$\color{blue}\text{c}\color{black}$. 
This stray field acts on the NiBi$_{3}$ layer and causes a significant detrimental effect on the superconducting state of the bilayers, particularly for the samples with thinner superconducting layers, resulting in the sharp reduction of the $T_c$. The influence of stray fields in Ni/Bi bilayers has been discussed in a previous study \cite{zhou2017magnetic}.

The primary factor contributing to the broadening of
the red and blue transition curves in Fig. \ref{fig:Tc_thickness}$\color{blue}\text{a}\color{black}$  probably comes from the stray field generated by the nearby ferromagnetic Ni. 
Although less likely, there might be another contribution to this effect: a crossover from a three-dimensional (3D) to a two-dimensional
(2D) behavior. This is expected to occur when the thickness of the superconducting layer becomes comparable
to or smaller than the coherent length. The change in dimensionality induces significant fluctuations in the order
parameter due to the influence of interface and surface
scattering, resulting in a strong decrease of the critical
temperature \cite{schneider1991dimensional}, similar to that shown in Fig. \ref{fig:Tc_thickness}$\color{blue}\text{b}\color{black}$. We will return to this point later in this manuscript.

\subsection{\label{sec:level}Superconducting properties}

\begin{figure}[H]
    \centering
    \includegraphics[height=11.5cm, width=8.5cm]{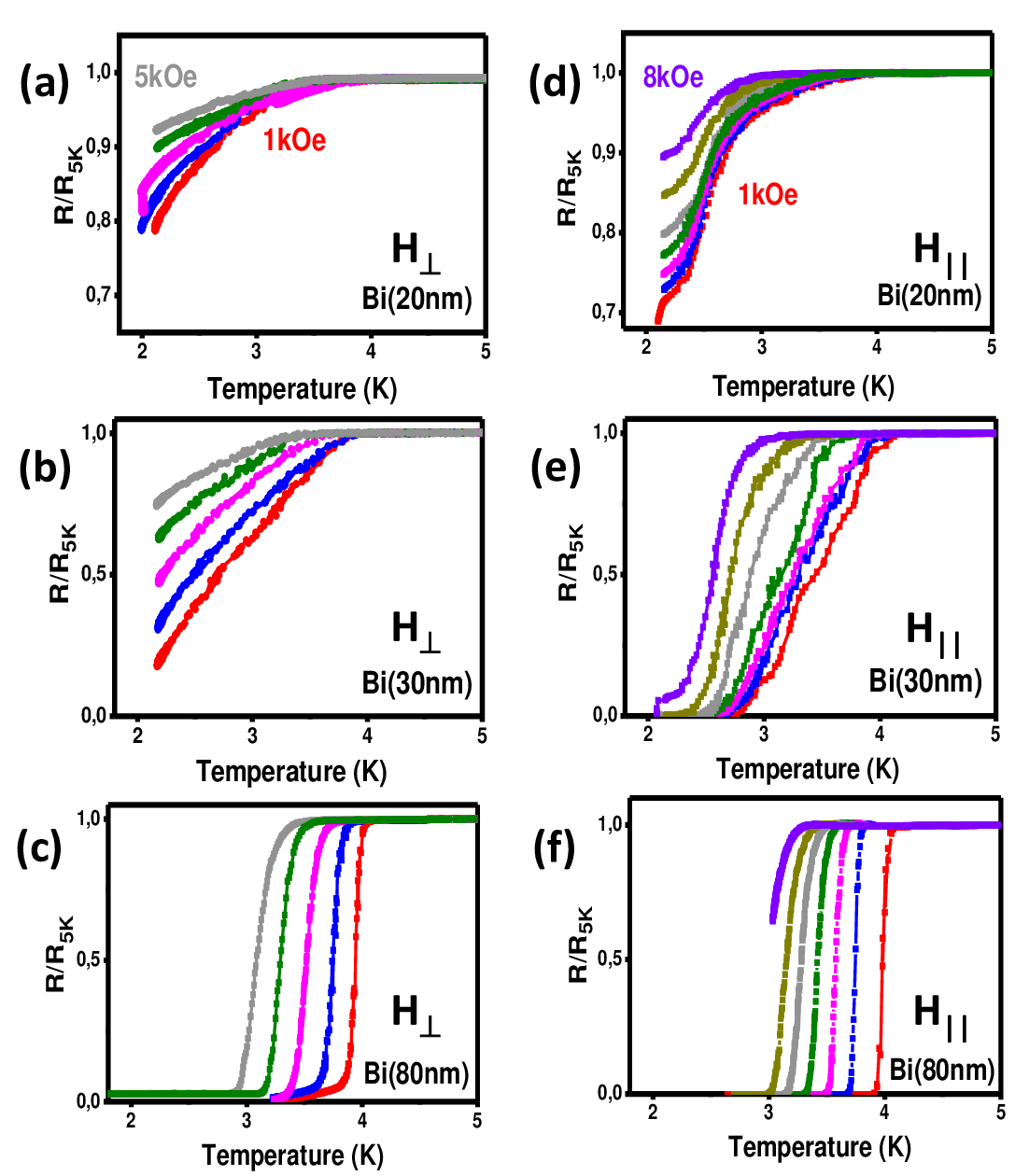}
    \caption{Normalized resistance in 5 K vs. temperature for perpendicular (a-c) and parallel (d-f) applied magnetic field orientation for Ni(8nm)/Bi($x$) samples with $x$ = 20, 30 and 80nm.
    The colors start from 1 kOe (red) up to 5 kOe (gray) for perpendicular and up to 8 kOe (purple) for parallel magnetic fields.}
    \label{fig:RxT_paral_perp}
\end{figure}

\begin{figure}[h!]
    \centering
    \includegraphics[height=11cm, width=9cm]{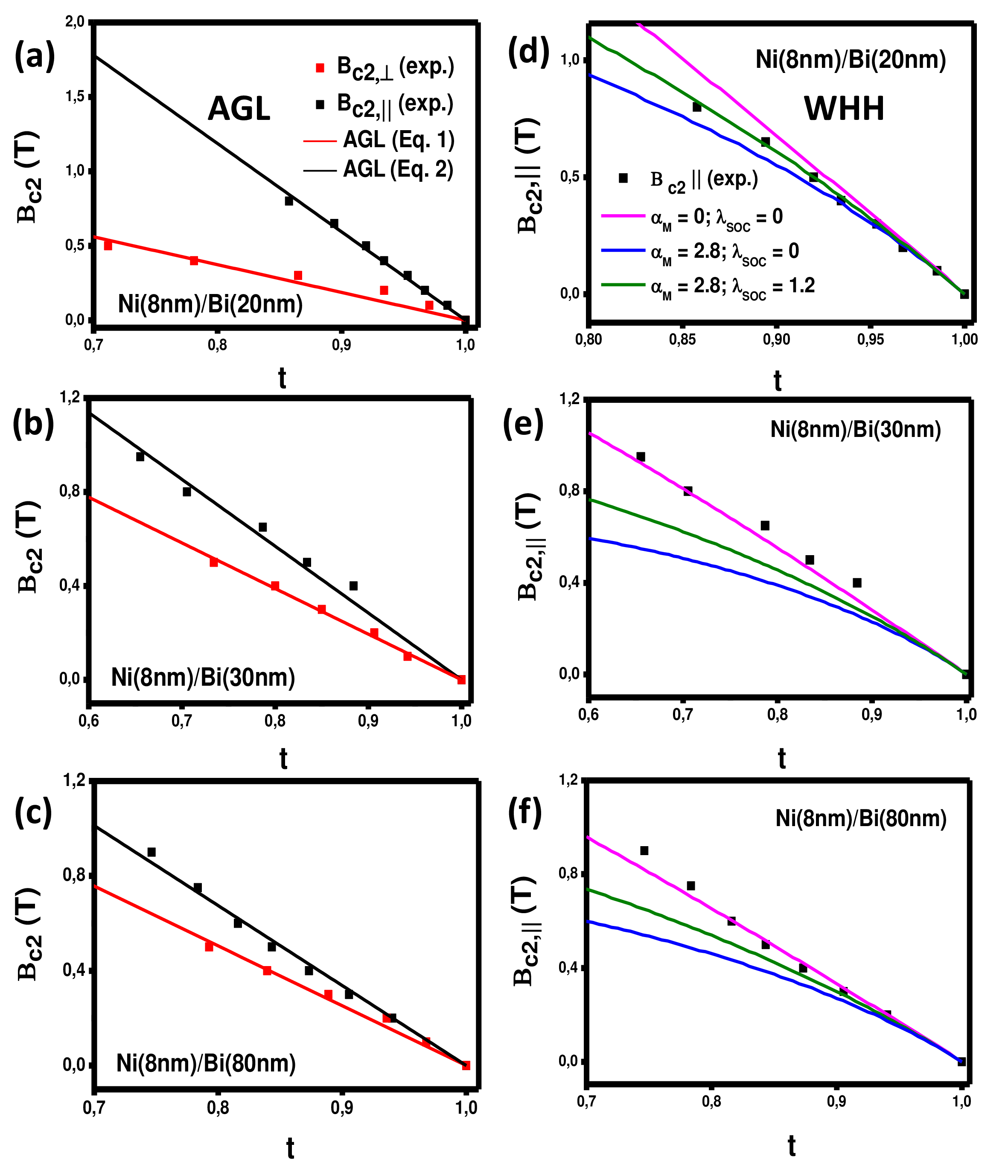}
    \caption{Upper critical field vs. temperature phase diagram analyzed within the scope of AGL theory in perpendicular (Eq. \ref{eqn: Bc2_perp}) and parallel 3D (Eq. \ref{eqn: Bc2_paral_3D}) fields for Ni(8nm)/Bi($x$) (a) $x$ = 20 nm; (b) $x$ = 30 nm and (c) $x$ = 80nm. Parallel upper critical field in function of reduced temperature within the scope of WHH model for Ni(8nm)/Bi($x$) (d) $x$ = 20 nm; (e) $x$ = 30 nm and (f) $x$ = 80 nm.}
    \label{fig:diagramadefase}
\end{figure}

In Fig. \ref{fig:RxT_paral_perp} we show the normalized resistance vs. temperature results measured at different magnetic fields for two field orientations, namely, perpendicular to the plane of the film $\color{blue}\text{(a-c)}\color{black}$ and parallel $\color{blue}\text{(d-f)}\color{black}$ to the electrical current. Applied fields are from 1-5 kOe in steps of one (colors varying from red to gray),
and 1-8 kOe  (colors varying from red to purple) for parallel and perpendicular field directions, respectively. 
Corrections for the demagnetizing effects are negligible in all cases. 
Increasing fields move the transition $T_c (B)$ towards lower temperatures as expected. 
Nevertheless, the width of the transition seems to increase much faster in samples with low Bi thickness, which can also be related to the stray field generated by the ferromagnetic Ni layer. 
From Fig. \ref{fig:RxT_paral_perp}, we extract the upper critical field $B_\mathrm{c2,\perp}$ and $B_\mathrm{c2,\parallel}$ as a function of the reduced temperature $t=T/T_c$. The obtained experimental points are shown in Fig. \ref{fig:diagramadefase}.

In the case of perpendicular magnetic fields the transition from the normal state to the superconducting vortex phase is described by the linear temperature dependence shown by the red points in Fig. \ref{fig:diagramadefase}$\color{blue}\text{a-c}\color{black}$ for all samples. 
Assuming that the orbital effect is the dominant limiting mechanism introduced by the applied magnetic field, we estimate the in-plane zero-temperature coherence length $\xi_\parallel(0)$ by fitting the red experimental data to the following expression from the anisotropic Ginzburg-Landau (AGL) model \cite{tinkham2004introduction}:
\begin{align}
    B_\mathrm{c2,\perp}(t)=\frac{\Phi_0}{2\pi\xi_\parallel^2(0)}(1-t), \label{eqn: Bc2_perp}
\end{align}
where $\Phi_0$ is the flux quantum. 
The extracted values for $\xi_\parallel(0)$ are listed in Tab. \ref{tab:coherence}. The values of $\xi_\parallel(0)$ are close to those reported by Vaughen \cite{vaughan2020origin} for a Ni/Bi bilayer of about 13.8 nm and slightly lower than those for a NiBi$_3$ single crystal \cite{zhu2012surface}, where $\xi_{a}$(0) = 18.1 nm and $\xi_{c}$(0) = 14.3 nm along crystalline a and c axis, respectively. 
Another relevant result is the weak dependence of $\xi_\parallel(0)$ with respect to the Bi thickness. Here, one would expect a reduction of $\xi_\parallel(0)$ (bulk-like) with the reduction of thickness \cite{pinto2018dimensional}, once thinner samples should present higher levels of disorder, as indeed shown by the residual resistivity ratio (RRR) also displayed in Tab. \ref{tab:coherence}. Consequently, we are lead to consider that disorder does not play a significant role in the properties of the superconducting state of our samples, at least those directly dependent on the coherence length.

The experimental data corresponding to in-plane magnetic fields are represented by the black points in Fig. \ref{fig:diagramadefase}$\color{blue}\text{a-c}\color{black}$. We fitted these results to the the 3D AGL model \cite{tinkham2004introduction}, which predicts the that the upper critical field for in-plane direction is given by

\begin{align}
B_\mathrm{c2,\parallel} = \frac{1.69\Phi_0}{2\pi\xi_\parallel(0)\xi_{\perp}(0)}\left(1 - t\right),\label{eqn: Bc2_paral_3D}
\end{align}
where the out-of-plane coherence length ($\xi_\perp$) is introduced by assuming the effective mass anisotropy approximation. By fitting the in-plane field data with the above expression (black line on Fig.  \ref{fig:diagramadefase}$\color{blue}\text{a-c}\color{black}$) we extracted the values of $\xi_\perp(0)$, which are also listed in Tab. \ref{tab:coherence}. 
  
For samples with 30 nm and 80 nm of Bi, the anisotropy factor defined as $\gamma=\xi_\perp(0)/\xi_\parallel(0)$ is close to 1, indicating an isotropic behavior similar to the bulk NiBi$_{3}$ type-II superconductor. 
On the other hand for the Ni(8nm)/Bi(20nm) specimen, the $\xi_\perp(0)$ is smaller than $\xi_\parallel(0)$, revealing the occurrence of an anisotropy in the coherence length that is characterized by the $\gamma=0.3$. 
At first glance, one might consider that
the reduction in $\xi_{\perp}(0)$  would be related to 2D effects due to the reduction of the NiBi$_{3}$ thickness. However, the 2D GL theory predicts that $B_\mathrm{c2,\parallel}$ varies with temperature as
$\sim (1-t)^{1/2}$
\cite{harper1968mixed}, which does not fit to our data.

\begin{table}[ht]
\centering
\caption{
Extracted coherence lengths within the framework of anisotropic Ginzburg-Landau theory, anisotropy factor $\gamma$ and residual resistivity ratio.}
\begin{tabular}{@{}lllll@{}}
\toprule
Sample & $\xi_\parallel(0)$ (nm) & $\xi_\perp(0)$ (nm) & $\gamma=\xi_\perp/\xi_\parallel$ & RRR \\ \midrule
Ni(8nm)/Bi(20nm) & 13.3$\pm$ 0.4 & 4.2$\pm$ 0.4 & 0.3 & 1.7\\
Ni(8nm)/Bi(30nm) & 13.1$\pm$ 0.2 & 15.1$\pm$ 0.6 & 1.2 & 1.8\\
Ni(8nm)/Bi(80nm) & 11.1$\pm$ 0.3 & 14.5$\pm$ 0.3 & 1.3 & 3.1\\\bottomrule
\end{tabular}
\label{tab:coherence}
\end{table}

Another possibility is the effect of a dimensional crossover. The enhancement of $B_\mathrm{c2,\parallel}$ provoked by the 3D-2D dimensional crossover usually occurs below a characteristic temperature known as T$^{*}$ \cite{schneider1991dimensional,tinkham2004introduction,chun1984dimensional,klemm1974upper}. 
In our case, however, the linear temperature dependence shown by the experimental data clearly implies the validity of the $T^{*} < T < T_c$ regime. For this reason, the parallel upper critical field should approach the 3D value of the ordinary type-II superconducting phase. 
Based on these arguments, we speculate that the reduction of $\xi_{\perp}(0)$ in thin samples may be related to an effect other than disorder and dimensional crossover.

The Ni/Bi system has been widely investigated due to the possibility of hosting unconventional superconductivity, where the spin-orbit coupling is supposed to play a key role. 
Assuming that the superconductivity in our bilayers stems from the NiBi$_3$ layer, this inevitably leads to a superconducting state that has a different environment close to the interfaces with Bi and Ni. 
We then speculate that the system studied here, the trilayer Ni/NiBi$_{3}$/Bi, in the limit of low Bi thickness, as we observed in the sample Ni(8nm)/Bi(20nm), could belong to the class of noncentrosymmetric superconductors. These materials experience inversion-breaking spin-orbit coupling effects enhancing the critical field.

To investigate the role of spin-orbit coupling in this system, we analyze the $B_\mathrm{c2,\parallel}$ with the Werthamer-Helfand-Hohenberg (WHH) model. Three calculated theoretical curves and the experimental points are shown in Figs. \ref{fig:diagramadefase}$\color{blue}\text{d-f}\color{black}$. This model incorporates both orbital and paramagnetic limiting mechanisms. 
The relative strength between these two effects is quantified by the Maki parameter \cite{maki1966effect}, defined as $\alpha_{M}$ = $\sqrt{2}$$B_\mathrm{c2,\text{orb.}}$(0)/$B_\mathrm{P}$(0), where $B_\mathrm{c2,\text{orb.}}$ and $B_\mathrm{P}$ are the orbital and paramagnetic limiting critical fields at zero temperature, respectively. For a single-band dirty-limit superconductor, the WHH model includes $\alpha_{M}$ and spin-orbit scattering ($\lambda_{SO}$). The overall upper critical field is obtained from the equation below \cite{werthamer1966temperature}:
\begin{eqnarray}
    \ln\left(\frac{1}{t}\right) = \sum_{v = -\infty}^{\infty} \bigg(\frac{1}{|2v + 1|} -\bigg[|2v + 1| + \frac{\bar{h}}{t}\nonumber + \frac{(\alpha_{M}\bar{h}/t)^{2}}{|2v + 1| + (\bar{h} + \lambda_{SO})/t}\bigg]\bigg), \label{eqn:whh}
\end{eqnarray}
where $\bar{h}$ = $\left(\frac{4}{\pi^{2}}\frac{B_\mathrm{c2}(t)}{|dB_\mathrm{c2}(t)/dt|}_{t = 1}\right)$.

Since the AGL model includes purely orbital effects, we considered $B_\mathrm{c2,\text{orb.}}$(0) as given by the slope obtained from fits with Eq. (\ref{eqn: Bc2_paral_3D}) near $T_c$, while $B_{P}(0) = 1.84T_c$ is established in the literature for BCS superconductors \cite{clogston1962upper}. Thus, for Ni(8 nm)/Bi(20 nm), Ni(8 nm)/Bi(30 nm) and Ni(8 nm)/Bi(80 nm) samples we roughly estimate $\alpha_{M}$ as 2.80, 0.34 and 0.38, respectively. 

Fig. \ref{fig:diagramadefase}$\color{blue}\text{e,f}\color{black}$ presents the WHH fit for the thicker samples, with Bi layers of 30 and 80 nm. Notice that the best agreement to the experiments occurs for both parameters $\alpha_{M}$ and $\lambda_{SO}$ equal to zero (magenta curve). 
This situation corresponds to $B_\mathrm{c2,\text{orb}}(0)\ll B_\mathrm{\text{P}}(0)$ indicating a field-induced suppression of superconductivity due to orbital effects, as expected for ordinary type-II superconductors. 
The case of Ni(8nm)/Bi(20nm) sample is more interesting, once the experimental points can be well described by introducing $\alpha_{M} = 2.8$ and $\lambda_{SO} = 1.2$, see Fig. \ref{fig:diagramadefase}$\color{blue}\text{d}\color{black}$. It is essential to emphasize that our plots do not guarantee the existence of spin-orbit coupling. 
However, it can be considered as evidence that, at low thicknesses, the manifestation of spin-orbit effects can potentially account for the changes observed in the superconducting properties of the Ni/Bi system.

The origin of SC in the Ni/Bi bilayers is still a topic of debate. 
Whether it occurs at the Bi layer near the interface or whether it stems from bulk NiBi$_{3}$ formed at the interface remains a controversial subject. 
In this work, we find evidence in favor of the second hypothesis for thicker samples. 
In contrast, we point out the possibility
of interesting effects occurring when very thin layers of NiBi$_{3}$ are formed at the bilayer interface. 
In this limit, the lack of inversion symmetry and the low thickness of the interfacial superconducting compound
leads to a peculiar phenomenology such as that originated by an anti-symmetric spin-orbit coupling \cite{smidman2017superconductivity}. 
This adds to the claims favoring unconventional superconductivity reported so far in the Ni/Bi system \cite{gong2017time,gong2015possible,chauhan2019nodeless,wang2017anomalous} and gives support to theoretical models as the one in Ref. \cite{chao2019superconductivity}. 
Additionally, we notice that bulk NiBi$_{3}$ has been recently suggested as hosting a topological surface state \cite{adriano2023bulk}, which might be a potential ingredient to be considered in the physics of thin Ni/Bi bilayers.

\section{\label{sec:level4}Conclusions}

We investigated the structural properties
and the temperature dependence of the upper critical field
of polycrystalline Ni(8nm)/Bi($x$) bilayers with $x$ = 10, 20, 30, 40, 60 and 80 nm. 
XRD, RBS and VSM analysis indicated that the original bilayers actually form the trilayer structure Ni/NiBi$_{3}$/Bi. 
For samples with large Bi thickness, the structure becomes effectively NiBi$_{3}$/Bi. 
In the trilayer, the stray field due to the remaining layer of ferromagnetic Ni leads to a decrease in the critical temperature and to a broadening of the superconducting transition. 
The upper critical field was analyzed in the context of the AGL and WHH models. For larger Bi layer thicknesses, the coherence length is nearly isotropic ($\gamma = \xi_{\perp}(0)/\xi_{\parallel}(0) \approx 1$) and has values similar to that found for NiBi$_{3}$ single crystals. On the other hand, for low Bi thicknesses, an anisotropic behavior is observed ($\gamma$ = 0.32). In the framework of the WHH model, thicker samples are described by a null Maki parameter and spin-orbit coupling. In the smallest Bi thickness sample, the Maki parameter ($\alpha_{M}$ = 2.8) and the spin-orbit scattering ($\lambda_{SO}$ = 1.2) are shown to be relevant to fully describe the temperature dependence of the upper critical field. Therefore, our results favor a thickness-dependent superconducting state in the Ni/Bi bilayer system.

\section*{Acknowledgments}
The authors would like to acknowledge the funding agencies FAPERGS, CAPES and CNPq, in terms of the following grants: Pronex grant no. 16/0490-0 (Fapergs-CNPq)

\nocite{*}
\bibliographystyle{IEEEtran}
\bibliography{main}

\end{document}